\documentclass[pdftex,twocolumn,epjc3]{svjour3}
\usepackage{amsfonts}
\usepackage{amsmath}
\usepackage{amssymb,bbold}
\usepackage[normalem]{ulem}

\RequirePackage[T1]{fontenc}
\smartqed
\RequirePackage{graphicx}
\RequirePackage{mathptmx}
\RequirePackage{flushend}
\RequirePackage[numbers,sort&compress]{natbib}
\RequirePackage[colorlinks,citecolor=blue,urlcolor=blue,linkcolor=blue]{hyperref}
\journalname{Eur. Phys. J. C}
\RequirePackage{color}

\begin{document}

\title{Quantum dynamics of a spin-1/2 charged particle in the presence of
magnetic field with scalar and vector couplings}
\author{Luis B. Castro \thanksref{e1} \and Edilberto O. Silva \thanksref{e2}}
\thankstext{e1}{e-mail: lrb.castro@ufma.br}
\thankstext{e2}{e-mail:
edilbertoos@pq.cnpq.br}

\institute{Departamento de F\'{i}sica,
  Universidade Federal do Maranh\~{a}o,
  Campus Universit\'{a}rio do Bacanga,
  65080-805 S\~{a}o Lu\'{i}s, Maranh\~{a}o, Brazil
         }

\journalname{Eur. Phys. J. C}
\date{Received: date / Accepted: date}
\maketitle

\begin{abstract}
The quantum dynamics of a spin-1/2 charged particle in the presence of
magnetic field is analyzed for the general case where scalar and vector
couplings are considered. The energy spectra are explicitly computed for
different physical situations, as well as their dependencies on the magnetic
field strength, spin projection parameter and vector and scalar coupling
constants.
\end{abstract}

\section{Introduction}

\label{sec:introduction}

The study of relativistic quantum systems under the influence of magnetic
field and scalar potentials has attracted attention of researchers in
various branches of physics. It is well known that these potentials can be
inserted into the Dirac equation
\begin{equation}
\left[ \beta \boldsymbol{\gamma }\cdot \mathbf{p}+\beta M\right] \psi \left(
\mathbf{r}\right) =E\psi \left( \mathbf{r}\right) ,  \label{dirac1}
\end{equation}%
through three usual substitutions, known as minimal, vector and scalar
couplings, whose representation is denoted, respectively, by%
\begin{eqnarray}
\mathbf{p} &\rightarrow &\mathbf{p}-e\mathbf{A,}  \label{ca} \\
E &\rightarrow &E-V\left( r\right) ,  \label{cb} \\
M &\rightarrow &M+S\left( r\right) .  \label{cc}
\end{eqnarray}%
a variety of relativistic and nonrelativistic effects can be studied. Those
couplings above differ in the manner how they are inserted into the Dirac
equation \cite{greiner.rqm.wf}. The minimal coupling (\ref{ca}) is useful
for studying the dynamics of a spin-1/2 charged particle in a magnetic
field. For example, using this model, we can study Landau levels \cite%
{Book.1981.Landau}, Aharonov-Bohm effect \cite{PR.1959.115.485}, Hall effect
\cite{AJM.1879.2.287}, and other effects associated with magnetic field.

It is known that the prescription (\ref{cb}) acts differently on electron
and positron states, respectively, and the eigenvalue spectrum of the
particle is not symmetric. In this case, bound states exist for only one of
the two kinds of particles. In other words, we can say that, for vector
coupling, the potential couples to the charge. In the context of the Dirac
equation, this coupling has been used, for example, to study the influence
of a harmonic oscillator on the Aharonov-Casher problem \cite%
{EPJC.2013.73.2402}, the Aharonov Bohm effect for a spin-1/2 particle in the
case that a $1/r$ potential is present \cite{AoP.1996.251.45}, effects of
nongauge potentials on the spin-1/2 Aharonoh-Bohm problem \cite%
{PRD.1993.48.5935}, quasiclassical theory of the Dirac equation with
applications in the physics of heavy-light mesons \cite{PRD.2011.83.076003}
and confining potentials with pure vector coupling \cite{PRL.2008.101.190401}%
. In the Schr\"{o}dinger theory, It also has important applications, such
as, in the dynamics of an electron in a two-dimensional quantum ring \cite%
{PRB.1996.53.6947,JMP.2012.53.023514}, quantum particles constrained to move
on a conical surface \cite{JMP.2012.53.122106} and effect of singular
potentials on the harmonic oscillator \cite{AoP.2010.325.2529}.

In the case of the scalar coupling (\ref{cc}), it is added to the mass term
of the Dirac equation and, therefore, it can be interpreted as an effective,
position-dependent mass and, furthermore, it also acts equally on particles
and antiparticles. This coupling has been used, for example, to obtain an
exact solution of the Dirac equation for a charged particle with
position-dependent mass in the Coulomb field \cite{PLA.2004.322.72}, to
study the relativistic quantum dynamics of a charged particle in cosmic
string spacetime \cite{EPJC.2012.72.2051,CQG.1996.13.L41}, scattering of a
fermion in the background of a smooth step potential with a general mixing
of vector and scalar Lorentz structures with the scalar coupling stronger
than or equal to the vector coupling \cite{AoP.2014.346.164}, \ inclusion of
the generalized Hulth\'{e}n potential in the case of the smooth step mass
distribution \cite{PLA.2006.352.478} and extension of PT-symmetric quantum
mechanics \cite{AoP.2008.323.566}. The coupling (\ref{cc}) also has
important applications in nonrelativistic quantum mechanics. The Schr\"{o}%
dinger equation with a position-dependent mass has attracted a lot of
attentions due to a wide range of applications in various areas of material
science and condensed matter physics. For example, to study the dynamics of
an one-dimensional harmonic oscillator \cite{CTP.2014.62.790}, derivation of
the Shannon entropy for a particle with a nonuniform solitonic mass density
\cite{AoP.2014.348.153}, context of displacement operator for quantum
systems \cite{PRA.2012.85.034102,PRA.2014.89.049904}, use of instantaneous
Galilean invariance to derive the expression for the Hamiltonian of an
electron \cite{PRA.1995.52.1845}, determination of some potential functions
for exactly solvable nonrelativistic problems \cite{PRA.2002.66.042116} and
Hermitian, rotationally invariant one-band Schr\"{o}dinger Hamiltonian \cite%
{PRB.1999.59.7596}.

The case in which the couplings are composed by a vector (\ref{cb}) and a
scalar (\ref{cc}) potentials, with $S=V$ ($S=-V$), are usually pointed out
as necessary condition for occurrence of exact spin (pseudospin) symmetry.
It is known that the spin and pseudospin symmetries are SU(2) symmetries of
a Dirac Hamiltonian with vector and scalar potentials. The pseudospin
symmetry was introduced in nuclear physics many years ago \cite%
{PLB.1969.30.517,NPA.1969.137.129} to account for the degeneracies of
orbital in single-particle spectra. Also, it is known that the spin symmetry
occurs in the spectrum of a meson with one heavy quark \cite{PRL.2001.86.204}
and anti-nucleon bound in a nucleus \cite{PR.1999.315.231}, and the
pseudospin symmetry occurs in the spectrum of nuclei \cite{PRL.1997.78.436}.

In this work, we study the quantum dynamics of a spin-1/2 charged particle
in the presence of magnetic field with scalar and vector couplings. This
system has been considered in Ref. \cite{AoP.2014.341.153}. The difference
between our approach and the previous one is that, here, we solve the
problem in a rigorous way taking into account other questions which have not
been analyzed by the authors. For example, absence of the term which depends
explicitly on the spin in the equation of motion. Since we are considering
the dynamics of a particle with spin, such a term can not be neglected in
equation of motion \cite{PRL.1990.64.503}. Moreover, as the authors make a
connection with the Aharonov-Bohm problem, the presence of this term has
important implications on the physical quantities of interest, such as
energy eigenvalues, scattering matrix and phase shift (see Ref. \cite%
{AoP.2013.339.510} for more details). By taking into account the term that
depends explicitly on the spin into Pauli equation, we address the system in
connection with the spin-1/2 Aharonov-Bohm problem \cite{IJMPA.1991.6.3119}
and analyze the following questions: (a) the existence of isolated solutions
to the first order equation Dirac and (b) general dynamics in all space,
including the $r=0$ region. We use the self-adjoint extension method to
determine the most relevant physical quantities, such as energy spectrum and
wave functions by applying boundary conditions allowed by the system.

The paper is organized as follows. In section \ref{sec2}, we consider the
Dirac equation in $(2+1)$ dimensions with minimal, scalar and vector
couplings, and derive the set of first order differential equations. These
equations are useful to investigate possible isolated solutions to the
problem. In section \ref{sec3}, we solve the first order Dirac equation in
connection with the Aharanov-Bohm problem and scalar and vector couplings.
We found that, for certain values assumed by the physical parameters of the
system, isolated solutions exist and discuss the limits of validity of them.
In section \ref{sec4}, we derive the Pauli equation and study the dynamics
of the system taking into account exact symmetry spin and pseudospin limits.
In section \ref{sec5}, we briefly discuss some concepts of the self-adjoint
extension method and specify the boundary conditions at the origin which
will be used. In section \ref{sec6}, expressions for the energy eigenvalues
and wave functions are determined for both symmetry limits and compare them
with the results of Ref. \cite{AoP.2014.341.153}. We verify that the
presence of the spin element in the equation of motion introduces a
correction in the expressions for the bound state energy eigenvalues. In
section \ref{sec7}, we present our concluding remarks.

\section{Equation of motion}

\label{sec2}

We begin with the Dirac equation (\ref{dirac1}) in $(2+1)$ dimensions in
polar coordinates ($\hbar=c=1$)
\begin{equation}
\left\{ \beta \boldsymbol{\gamma }\cdot \mathbf{\pi }+\beta \left[ M+S\left(
r\right) \right] \right\} \psi \left( \mathbf{r}\right) =\left[ E-V\left(
r\right) \right] \psi \left( \mathbf{r}\right) ,  \label{dirac2}
\end{equation}%
where $\mathbf{\pi }=\left( \pi _{r},\pi _{\varphi }\right) =(-i\partial
_{r},-i\partial _{\varphi }/r-eA_{\varphi })$, $\mathbf{r}=(r,\varphi )$ and
$\psi $ is a two-component spinor. The $\boldsymbol{\gamma }$ matrices in
Eq. (\ref{dirac2}) are given in terms of the Pauli matrices as \cite%
{EPJC.2014.74.3187}
\begin{align}
\beta \gamma ^{r}& =\sigma _{1}\cos \varphi +s\sigma _{2}\sin \varphi
=\left(
\begin{array}{cc}
0 & e^{-is\varphi } \\
e^{+is\varphi } & 0%
\end{array}%
\right),  \label{sgimar} \\
\beta \gamma ^{\varphi }& =-\sigma _{1}\cos \varphi +s\sigma _{2}\sin
\varphi =\left(
\begin{array}{cc}
0 & -ise^{-is\varphi } \\
ise^{+is\varphi } & 0%
\end{array}%
\right) ,  \label{sgimaphi} \\
\beta & =\sigma _{3}=\left(
\begin{array}{cc}
1 & 0 \\
0 & -1%
\end{array}%
\right) ,  \label{sgimaz}
\end{align}%
where $s$ is twice the spin value, with $s=+1$ for spin \textquotedblleft
up\textquotedblright\ and $s=-1$ for spin \textquotedblleft
down\textquotedblright . Equation (\ref{dirac2}) can be written more
explicitly as%
\begin{eqnarray}
e^{-is\varphi }\left[ \pi _{r}-is\pi _{\varphi }\right] \psi _{2} &=&\left[
E-M-\Sigma \left( r\right) \right] \psi _{1},  \label{dirac3a} \\
e^{+is\varphi }\left[ \pi _{r}+is\pi _{\varphi }\right] \psi _{1} &=&\left[
E+M-\Delta \left( r\right) \right] \psi _{2},  \label{dirac3b}
\end{eqnarray}%
where $\Sigma \left( r\right) =V\left( r\right) +S\left( r\right) $ and $%
\Delta \left( r\right) =V\left( r\right) -S\left( r\right) $.

If one adopts the following decomposition
\begin{equation}
\left(
\begin{array}{c}
\psi _{1} \\
\psi _{2}%
\end{array}%
\right) =\left(
\begin{array}{c}
\sum\limits_{m}f_{m}(r)\;e^{im\varphi } \\
i\sum\limits_{m}g_{m}(r)\;e^{i(m+s)\varphi }%
\end{array}%
\right) ,  \label{ansatz}
\end{equation}%
with $m+1/2=\pm 1/2,\pm 3/2,\ldots $, with $m\in \mathbb{Z}$, and inserting
this into Eqs. (\ref{dirac3a}) and (\ref{dirac3b}), we obtain%
\begin{eqnarray}
\left[ \frac{d}{dr}+\frac{s\left( m+s\right) }{r}-esA_{\varphi }\right]
g\left( r\right) &=&\left[ E-M-\Sigma \left( r\right) \right] f\left(
r\right) ,  \label{dirac4} \\
\left[ -\frac{d}{dr}+\frac{sm}{r}-esA_{\varphi }\right] f\left( r\right) &=&%
\left[ E+M-\Delta \left( r\right) \right] g\left( r\right) .  \label{dirac5}
\end{eqnarray}

Note that the above equations are coupled. However, if $\Sigma \left(
r\right) $ or $\Delta \left( r\right) $ is made zero in any of the
equations, we can uncouple them easily. We will see below that this result
in important physical consequences to the physical system in question.

\section{Isolated solutions for the Dirac equation of motion}

\label{sec3}

In this section, we investigate the existence of isolated solutions in the
quantum motion of a fermionic massive charged particle in $(2+1)$
dimensions. This is accomplished by considering the particle at rest, i.e., $%
E=\pm M$, directly in the first order equations in Eqs. (\ref{dirac4}) and (%
\ref{dirac5}). Such a solution is known to be excluded from the
Sturm-Liouville problem, and have been investigated under diverse
perspectives in the latest years \cite%
{EPL.2014.108.30003,AoP.2013.338.278,PS.2008.77.045007,IJMPE.2007.16.2998,IJMPE.2007.16.3002, JPA.2007.40.263,PS.2007.75.170,PLA.2006.351.379}%
. We are seeking for bound-state solutions subjected to the normalization
condition
\begin{equation}
\int_{0}^{\infty }\left( |f_{m}(r)|^{2}+|g_{m}(r)|^{2}\right) rdr=1.
\label{norm}
\end{equation}

In order to determine the isolated bound-state solutions, we consider $%
\Sigma \left( r\right) =0$ in Eq. (\ref{dirac4}), so that, for $E=M$, we can
write
\begin{eqnarray}
&&\left[ \frac{d}{dr}+\frac{s(m+s)}{r}-seA_{\varphi }\right] g_{m}(r)=0,
\label{diracme} \\
&&\left[ -\frac{d}{dr}+\frac{sm}{r}-seA_{\varphi }\right] f_{m}(r)=2\left(
M-V\left( r\right) \right) g_{m}(r),  \label{diracma}
\end{eqnarray}%
whose general solutions are
\begin{eqnarray}
g_{m}(r) &=&a_{+}r^{-s\left( m+s\right) }e^{se\int A_{\varphi }dr},
\label{Mma} \\
f_{m}(r) &=&\left[ b_{+}-a_{+}I(r)\right] r^{sm}e^{-se\int A_{\varphi }dr},
\label{Mmb}
\end{eqnarray}%
where $a_{+}$ and $b_{+}$ are constants, and $I(r)$ is given by
\begin{equation}
I(r)=\int dr\left[ 2M-2V\left( r\right) \right] e^{2se\int A_{\varphi
}\left( r\right) dr},
\end{equation}%
which, for a given $V\left( r\right) $ and $A_{\varphi }\left( r\right) $,
it can be expressed in terms of the upper incomplete Gamma function \cite%
{Book.1972.Abramowitz}
\begin{equation}
\Gamma (a,x)=\int_{x}^{\infty }t^{a-1}e^{-t}dt,\qquad \Re (a)>0.
\label{eq:incgamma}
\end{equation}%
Let us now analyze solutions for $E=-M$, and consider $\Delta \left(
r\right) =0$ in Eq. (\ref{dirac5}). For this case, we write
\begin{align}
& \left[ \frac{d}{dr}+\frac{s\left( m+s\right) }{r}-esA_{\varphi }\left(
r\right) \right] g\left( r\right)=-2\left[ M+V\left( r\right) \right]
f\left( r\right), \\
& \left[ -\frac{d}{dr}+\frac{sm}{r}-esA_{\varphi }\left( r\right) \right]
f\left( r\right) =0,
\end{align}%
whose general solution is
\begin{eqnarray}
f_{m}(r) &=&a_{-}r^{sm}e^{-se\int A_{\varphi }dr},  \label{fa} \\
g_{m}(r) &=&\left[ b_{-}-a_{-}H(r)\right] r^{-s\left( m+s\right) }e^{se\int
A_{\varphi }dr},  \label{ga}
\end{eqnarray}%
where
\begin{equation}
H(r)=\int dr\left[ 2M-2V\left( r\right) \right] e^{-2se\int A_{\varphi }dr}.
\end{equation}

Now, lets us consider the particular case where the particle moves in a
constant magnetic field and in the presence of Aharonov-Bohm effect. The
vector potential in the Coulomb gauge is
\begin{equation}
\mathbf{A=A}_{1}+\mathbf{A}_{2},  \label{vectorA}
\end{equation}%
with
\begin{equation}
\mathbf{A}_{1}=\frac{B_{0}r}{2}\boldsymbol{\hat{\varphi}},~~~\mathbf{A}_{2}=%
\frac{\phi }{r}\boldsymbol{\hat{\varphi}},  \label{potential}
\end{equation}%
where $B_{0}$ is the magnetic field magnitude and $\phi $ is the flux
parameter. The potentials in Eq. (\ref{vectorA}) both provide one magnetic
field perpendicular to the plane $\left( r,\varphi \right) $, namely%
\begin{equation}
\mathbf{B}=\mathbf{B}_{1}+\mathbf{B}_{2},
\end{equation}%
with%
\begin{eqnarray}
\mathbf{B}_{1} &=&\mathbf{\nabla }\times \mathbf{A}_{1}=B_{0}\mathbf{\hat{z}}%
,  \label{fieldB1} \\
\mathbf{B}_{2} &=&\mathbf{\nabla }\times \mathbf{A}_{2}=\phi \frac{\delta (r)%
}{r}\mathbf{\hat{z}},  \label{fieldB2}
\end{eqnarray}%
where $\mathbf{B}_{1}$is an external magnetic field and $\mathbf{B}_{2}$ is
the magnetic field due to a solenoid. If the solenoid is extremely long, the
field inside is uniform, and the field outside is zero. However, in a
general dynamics, the particle is allowed to access the $r=0$ region. In
this region, the magnetic field is non-null. If the radius of the solenoid
is $r_{0}\approx 0$, then the relevant magnetic field is $\mathbf{B}_{2}\sim
\delta (r)$ as in Eq. (\ref{fieldB2}). This situation has not been
accomplished in Ref. \cite{AoP.2014.341.153}, which is crucial to give
meaning to the term that explicitly depends of the spin, namely, the Pauli
term that appearing in the second order differential equation. This issue
will be considered later when we treat solutions for the case $E\neq \pm M$.

Using Eqs. (\ref{fieldB1}) and (\ref{fieldB2}), we have
\begin{equation}
\int A_{\varphi }dr=\frac{B_{0}r^{2}}{4}+\phi \ln r.  \label{int}
\end{equation}%
If $\phi >0$ and $B_{0}>0$, for $E=M$, we have bound-state solutions only in
the following cases:
\begin{align}
\left(
\begin{array}{c}
f_{m}(r) \\
g_{m}(r)%
\end{array}%
\right) & =\left(
\begin{array}{c}
1 \\
0%
\end{array}%
\right) b_{+}r^{m-\lambda }e^{-\delta r^{2}},~\Bigg\{%
\begin{tabular}{l}
\hspace{-0.3cm} $s=+1,$ \\
\hspace{-0.3cm} $a_{+}=0,$%
\end{tabular}%
,  \label{ss1} \\
\left(
\begin{array}{c}
f_{m}(r) \\
g_{m}(r)%
\end{array}%
\right) & =\left(
\begin{array}{c}
0 \\
1%
\end{array}%
\right) a_{+}r^{m-\lambda -1}e^{-\delta r^{2}},~\Bigg\{%
\begin{tabular}{l}
\hspace{-0.3cm} $s=-1,$ \\
\hspace{-0.3cm} $b_{+}=M=V=0,$%
\end{tabular}
\label{ss2}
\end{align}%
and for $E=-M$,
\begin{eqnarray}
\left(
\begin{array}{c}
f_{m}(r) \\
g_{m}(r)%
\end{array}%
\right) &=&\left(
\begin{array}{c}
0 \\
1%
\end{array}%
\right) b_{-}r^{m-\lambda -1}e^{-\delta r^{2}},~\Bigg\{%
\begin{tabular}{l}
\hspace{-0.2cm}$s=-1,$ \\
\hspace{-0.2cm}$a_{-}=0,$%
\end{tabular}
\label{ss3} \\
\left(
\begin{array}{c}
f_{m}(r) \\
g_{m}(r)%
\end{array}%
\right) &=&\left(
\begin{array}{c}
1 \\
0%
\end{array}%
\right) a_{-}r^{m-\lambda }e^{-\delta r^{2}},~\Bigg\{%
\begin{tabular}{l}
\hspace{-0.3cm} $s=+1,$ \\
\hspace{-0.3cm} $b_{+}=M=V=0,$%
\end{tabular}
\label{ss4}
\end{eqnarray}%
where $\delta =eB_{0}/4$ and $\lambda =e\phi $. Note that the above results
are independent of the values of $s$, $m$ and $\lambda $ to ensure a bound
state. This is because the function $e^{-\delta r^{2}}$ predominates over
the polynomials $r^{m-\lambda }$ and $r^{m-\lambda -1}$. If we consider $%
B_{0}=V\left( r\right) =0$, which leads to the usual Aharonov-Bohm effect,
the solutions for $E=M$ reads
\begin{eqnarray}
g_{m}(r) &=&a_{+}r^{s\left[ \lambda -\left( m+s\right) \right] },\text{ }
\label{cs1} \\
f_{m}(r) &=&\left[ b_{+}-a_{+}\tilde{I}(r)\right] r^{s\left( m-\lambda
\right) },  \label{cs2}
\end{eqnarray}%
where
\begin{equation}
\tilde{I}(r)=\frac{2M}{2s\lambda +1}r^{2s\lambda +1}\,.
\end{equation}%
For $E=-M$, we get
\begin{align}
f_{m}(r)& =a_{-}r^{s\left( m-\lambda \right) },  \label{cs3} \\
g_{m}(r)& =\left[ b_{-}-a_{-}\tilde{H}(r)\right] r^{s\left[ \lambda -\left(
m+s\right) \right]},  \label{cs4}
\end{align}%
where
\begin{equation}
\tilde{H}(r)=\frac{2M}{-2s\lambda +1}r^{-2s\lambda +1},  \label{inth}
\end{equation}%
Unlike from cases of Eqs. (\ref{ss1}), (\ref{ss2}), (\ref{ss3}) and (\ref%
{ss4}), if we impose that $B_{0}$ and $V\left( r\right) $ are zero, there no
exist bound-state solutions of square-integrable. In other words, for any
values of $s$, $m$ and $\lambda$ in Eqs. (\ref{cs1})-(\ref{inth}), the
integral (\ref{norm}) diverges.

\section{Equation of motion and analysis of symmetries}

\label{sec4}

In this section, we investigate the dynamics for $E\neq \pm M$. For this, we
choose to work with Eq. (\ref{dirac2}) in its quadratic form. After
application of the operator%
\begin{equation}
\beta \left[ \left( M+S\left( r\right) \right) +\beta \left( E-V\left(
r\right) \right) +\boldsymbol{\gamma }\cdot \boldsymbol{\pi }\right] ,
\end{equation}%
we get%
\begin{eqnarray}
&&\left\{ \mathbf{p}^{2}-2e\left[ \left( \mathbf{A}_{1}+\mathbf{A}%
_{2}\right) \cdot \mathbf{p}\right] +e^{2}\left( \mathbf{A}_{1}+\mathbf{A}%
_{2}\right) ^{2}\right\} \psi \left( \mathbf{r}\right)  \notag \\
&&+\left\{ \left[ M+S\left( r\right) \right] ^{2}-\left[ E-V\left( r\right) %
\right] ^{2}-es\boldsymbol{\sigma }\cdot \left( \mathbf{B}_{1}+\mathbf{B}%
_{2}\right) \right\} \psi \left( \mathbf{r}\right)  \notag \\
&&-\left( \frac{\partial S\left( r\right) }{\partial r}\sigma _{2}+i\frac{%
\partial V\left( r\right) }{\partial r}\sigma _{1}\right) \psi \left(
\mathbf{r}\right) =0.  \label{diracB}
\end{eqnarray}
In this stage, it is worthwhile to mention that the Eq.~(\ref{diracB}) is
the correct quadratic form of the Dirac equation with minimal, vector and
scalar couplings, because the Pauli term is considered.

\subsection{Exact spin symmetry limit: $S=V$}

The condition for establishing the exact symmetry boundary implies that the
solution is of the form%
\begin{equation}
\psi _{1}=\sum\limits_{m}f_{m}(r)\;e^{im\varphi }.  \label{solsm}
\end{equation}%
So, by making $S=V$ (or equivalently $\Delta \left( r\right) =0$ \cite%
{PRC.1999.59.154,PRC.1998.58.R628}) in Eq. (\ref{dirac3b}) and using the
solution (\ref{solsm}) in Eq. (\ref{diracB}), the equation for $f_{m}\left(
r\right) $ is found to be%
\begin{eqnarray}
&&\left[ -\frac{d^{2}}{dr^{2}}-\frac{1}{r}\frac{d}{dr}+\frac{m^{2}}{r^{2}}%
-2e\left( \frac{B_{0}r}{2}+\frac{\phi }{r}\right) \frac{m}{r}+\frac{%
e^{2}B_{0}^{2}r^{2}}{4}\right] f_{m}\left( r\right)  \notag \\
&&+\left[ \frac{e^{2}\phi ^{2}}{r^{2}}+e^{2}B_{0}\phi -es\left( B_{0}+\phi
\frac{\delta (r)}{r}\right) \right] f_{m}\left( r\right)  \notag \\
&&+\left[ M^{2}-E^{2}+2\left( E+M\right) V\right] f_{m}\left( r\right) =0.
\label{dra}
\end{eqnarray}%
Assuming $V\left( r\right) $ as in Ref. \cite{AoP.2014.341.153}, i.e., of
the form
\begin{equation}
V\left( r\right) =a\,r^{2}+\frac{b}{r^{2}},  \label{vr}
\end{equation}%
Eq. (\ref{dra}) becomes%
\begin{equation}
Hf_{m}\left( r\right) =k^{2}f_{m}\left( r\right) ,  \label{eigen}
\end{equation}%
with%
\begin{equation}
H=H_{0}-es\phi \frac{\delta (r)}{r},  \label{hfull}
\end{equation}%
\begin{equation}
H_{0}=-\frac{d^{2}}{dr^{2}}-\frac{1}{r}\frac{d}{dr}+\frac{\nu ^{2}}{r^{2}}%
+\eta ^{2}\text{ }r^{2},  \label{hzero}
\end{equation}%
where%
\begin{eqnarray}
\nu ^{2} &=&\left( m-e\phi \right) ^{2}+2b\left( E+M\right) , \\
\eta ^{2} &=&\frac{e^{2}B_{0}^{2}}{4}+2a\left( E+M\right) , \\
k^{2} &=&meB_{0}-e^{2}B_{0}\phi +esB_{0}+\left( E^{2}-M^{2}\right) .
\end{eqnarray}
As pointed out in Ref. \cite{AoP.2014.341.153}, the potential $V(r)$ in Eq. (%
\ref{vr}) describes an anharmonic oscillator. This model is a particular
case of a proposed in Ref. \cite{PRB.1996.53.6947} to study the Landau
quantization and the Aharonov-Bohm effect in a two-dimensional ring as an
exactly soluble model. The model considered in Ref. \cite{PRB.1996.53.6947}
has an advantage because, besides the model considered here, it also
describes other physical systems, such as a one-dimensional ring, a straight
2D wire, a single quantum dot for and an isolated antidot.

\subsection{Exact pseudospin symmetry limit: $S=-V$}

In this case, the condition for establishing the exact pseudospin symmetry
limit implies that the resolution is related to the down component of the
spinor in Eq. (\ref{ansatz}), namely
\begin{equation}
\psi _{2}=i\sum\limits_{m}g_{m}(r)\;e^{i(m+s)\varphi }.  \label{solpsm}
\end{equation}
By making $S=-V$ (or equivalently $\Sigma \left( r\right) =0$ in Eq. (\ref%
{dirac3a}) and again using Eq. (\ref{solpsm}) in Eq. (\ref{diracB}), the
equation for $g_{m}\left( r\right) $ can be found%
\begin{equation}
\tilde{H}g_{m}\left( r\right) =\tilde{k}^{2}g_{m}\left( r\right) ,
\end{equation}%
with%
\begin{equation}
\tilde{H}=\tilde{H}_{0}-es\phi \frac{\delta (r)}{r},
\end{equation}%
\begin{equation}
H_{0}=-\frac{d^{2}}{dr^{2}}-\frac{1}{r}\frac{d}{dr}+\frac{\tilde{\nu}^{2}}{%
r^{2}}+\tilde{\eta}^{2}\text{ }r^{2},
\end{equation}%
where%
\begin{eqnarray}
\tilde{\nu}^{2} &=&\left( m+s-e\phi \right) ^{2}+2b\left( E-M\right) , \\
\tilde{\eta}^{2} &=&\frac{e^{2}B_{0}^{2}}{4}+2a\left( E-M\right) , \\
k^{2} &=&\left( m+s\right) eB_{0}-e^{2}B_{0}\phi +esB_{0}-\left(
M^{2}-E^{2}\right) .
\end{eqnarray}

\section{Self-adjoint extension analysis}

\label{sec5}

In this section, we review some concepts on the self-adjoint extension
approach. An operator $\mathcal{O}$, with domain $\mathcal{D}({\mathcal{O}})$%
, is said to be self-adjoint if and only if $\mathcal{O}=\mathcal{O}%
^{\dagger }$ and $\mathcal{D}(\mathcal{O})=\mathcal{D}(\mathcal{O}^{\dagger
})$, $\mathcal{O}^{\dagger }$ being the adjoint of operator $\mathcal{O}$.
For smooth functions, $\xi \in C_{0}^{\infty }(\mathbb{R}^{2})$ with $\xi
(0)=0$, we should have $H\xi =H_{0}\xi $, and it is possible to interpret
the Hamiltonian $H_{0}$ (\ref{hzero}) as a self-adjoint extension of $%
H_{0}|_{C_{0}^{\infty }(\mathbb{R}^{2}/\{0\})}$ \cite%
{crll.1987.380.87,JMP.1998.39.47,LMP.1998.43.43}. The self-adjoint extension
approach consists, essentially, in extending the domain of $\mathcal{D}(%
\mathcal{O})$ in order to match $\mathcal{D}(\mathcal{O}^{\dagger })$. From
the theory of symmetric operators, it is a well-known fact that the
symmetric radial operator $H_{0}$ is essentially self-adjoint for $\nu \geq
1 $, while for $\nu <1$, it admits an one-parameter family of self-adjoint
extensions \cite{Book.1975.Reed.II}, $H_{0,\lambda _{m}}$, where $\lambda
_{m}$ is the self-adjoint extension parameter. To characterize this family,
we will use the approach in \cite{JMP.1985.26.2520,Book.2004.Albeverio},
which is based in a boundary conditions at the origin. All the self-adjoint
extensions $H_{0,\lambda _{m}}$ of $H_{0}$ are parameterized by the boundary
condition at the origin
\begin{equation}
\Psi _{0}=\lambda _{m}\Psi _{1},  \label{bc}
\end{equation}%
with
\begin{align}
\Psi _{0}={}& \lim_{r\rightarrow 0^{+}}r^{\nu }f_{m}(r), \\
\Psi _{1}={}& \lim_{r\rightarrow 0^{+}}\frac{1}{r^{\nu }}\left[
f_{m}(r)-\Psi _{0}\frac{1}{r^{\nu }}\right] ,
\end{align}%
where $\lambda _{m}\in \mathbb{R}$ is the self-adjoint extension parameter.
For $\lambda _{m}=0$, we have the free Hamiltonian (without the $\delta $
function) with regular wave functions at the origin, and for $\lambda
_{m}\neq 0$ the boundary condition in Eq. (\ref{bc}) permit an $r^{-\nu }$
singularity in the wave functions at the origin.

\section{The bound state energy and wave function}

\label{sec6}

In this section, we determine the energy spectrum by solving Eq. (\ref{eigen}%
). For $r\neq 0$, the equation for the component $f_{m}(r)$ can be
transformed by the variable change $\rho =\eta r^{2}$ resulting in
\begin{equation}
\rho f_{m}^{\prime \prime }(\rho )+f_{m}^{\prime }(\rho )-\left( \frac{{\nu }%
^{2}}{4\rho }+\frac{\rho }{4}-\frac{k^{2}}{4\eta }\right) f_{m}(\rho )=0,
\label{edofrho}
\end{equation}%
Due to the boundary condition in Eq. (\ref{bc}), we seek for regular and
irregular solutions for Eq. (\ref{edofrho}). Studying the asymptotic limits
of Eq. (\ref{edofrho}) leads us to the following regular ($+$) (irregular ($%
- $)) solution
\begin{equation}
f_{m}(\rho )=\rho ^{\pm \frac{\nu }{2}}\mathrm{e}^{-\frac{\rho}{2}}F(\rho ).
\label{frho}
\end{equation}%
With this, Eq. (\ref{edofrho}) is rewritten as
\begin{equation}
\rho F^{\prime \prime }(\rho )+\left( 1\pm \nu -\rho \right) F^{\prime
}(\rho )-\left( \frac{1\pm \nu }{2}-\frac{k^{2}}{4\gamma }\right) F(\rho )=0.
\label{edoMrHo}
\end{equation}%
Equation (\ref{edofrho}) is of the confluent hypergeometric equation type
\begin{equation}
zF^{\prime \prime }(z)+(b-z)F^{\prime }(z)-aF(z)=0.
\end{equation}%
In this manner, the general solution for Eq. (\ref{edofrho}) is
\begin{eqnarray}
f_{m}(r) &=&a_{m}\rho ^{\frac{\nu }{2}}\mathrm{e}^{-\frac{\rho }{2}%
}\;F\left( d_{+},1+\nu ,\rho \right)  \notag \\
&+&b_{m}\rho ^{-\frac{\nu }{2}}\mathrm{e}^{-\frac{\rho }{2}}\;F\left(
d_{-},1-\nu ,\rho \right) ,  \label{general_sol_2_HO}
\end{eqnarray}%
with
\begin{equation}
d_{\pm }=\frac{1\pm \nu }{2}-\frac{k^{2}}{4\eta }.
\end{equation}%
In Eq. (\ref{general_sol_2_HO}), $F(a,b,z)$ is the confluent hypergeometric
function of the first kind \cite{Book.1972.Abramowitz} and $a_{m}$ and $%
b_{m} $ are, respectively, the coefficients of the regular and irregular
solutions.

In this point, we apply the boundary condition in Eq. (\ref{bc}). Doing
this, one finds the following relation between the coefficients $a_{m}$ and $%
b_{m}$:
\begin{equation}
\lambda _{m}\eta ^{\nu }=\frac{b_{m}}{a_{m}}\left[ 1+\frac{\lambda _{m}k^{2}%
}{4(1-\nu )}\lim_{r\rightarrow 0^{+}}r^{2-2\nu }\right] .  \label{coef_rel_1}
\end{equation}%
We note that $\lim_{r\rightarrow 0^{+}}r^{2-2\nu }$ diverges if $\nu \geq 1$%
. This condition implies that $b_{m}$ must be zero if $\nu \geq 1$ and only
the regular solution contributes to $f_{m}(r)$. For $\nu <1$, when the
operator $H_{0}$ is not self-adjoint, there arises a contribution of the
irregular solution to $f_{m}(r)$ \cite%
{AoP.2013.339.510,JPG.2013.40.075007,AoP.2008.323.1280,TMP.2009.161.1503,EPJC.2014.74.2708}%
. In this manner, the contribution of the irregular solution for system wave
function steams from the fact that the operator $H_{0}$ is not self-adjoint.
\begin{figure}[tbp]
\includegraphics[width=1.0 \linewidth,angle=0]{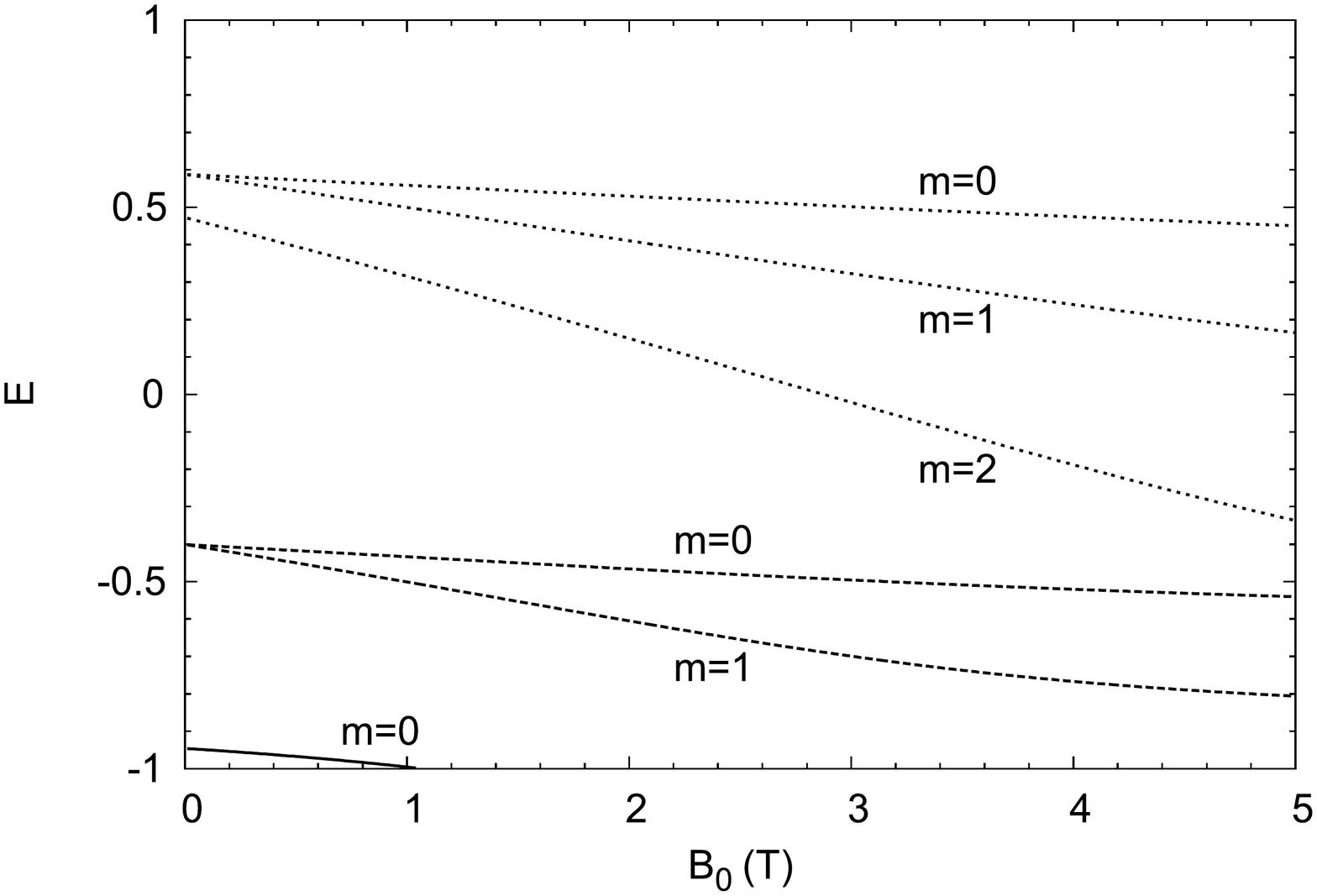}
\caption{Plots of the energy ($\Delta(r)=0$) as a function of the magnetic
field $B_{0}$ for $s=1$ and different values of $n$ and $m$: $n=0$ [solid
line], $n=1$ [dashed line] and $n=2$ [dotted line].}
\label{gra1}
\end{figure}
\begin{figure}[tbp]
\includegraphics[width=1.0 \linewidth,angle=0]{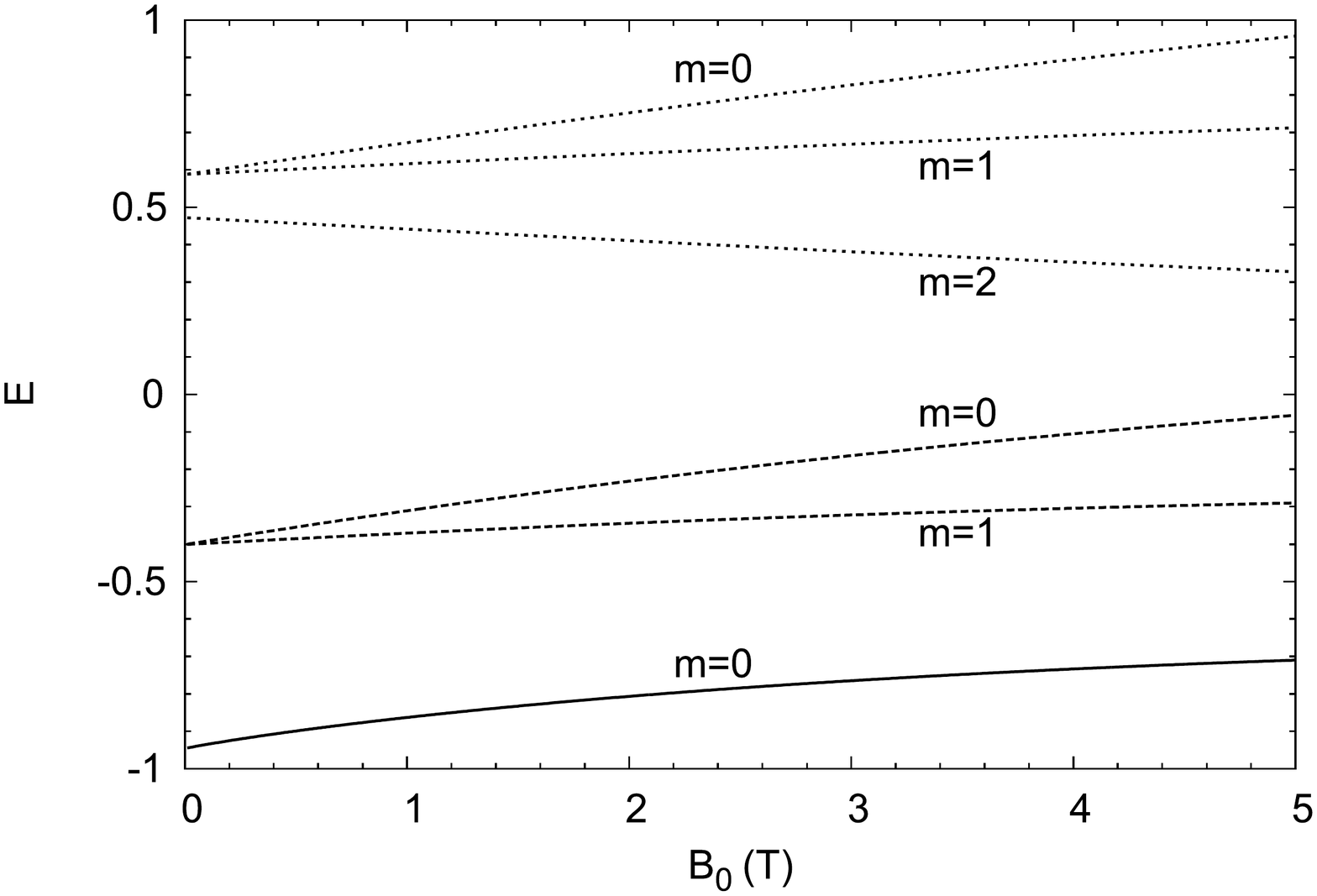}
\caption{Plots of the energy ($\Delta(r)=0$) as a function of the magnetic
field $B_{0}$ for $s=-1$ and different values of $n$ and $m$: $n=0$ [solid
line], $n=1$ [dashed line] and $n=2$ [dotted line].}
\label{gra2}
\end{figure}
For $f_{m}(r)$ be a bound state wave function, it must vanish at large
values of $r$, i.e., it must be normalizable. So, from the asymptotic
representation of the confluent hypergeometric function, the normalizability
condition is translated in
\begin{equation}
\frac{b_{m}}{a_{m}}=-\frac{\Gamma (1+\nu )}{\Gamma (1-\nu )}\frac{\Gamma
(d_{-})}{\Gamma (d_{+})}.  \label{coef_rel_2_DO}
\end{equation}%
From Eq. (\ref{coef_rel_1}), for $\nu <1$, we have
\begin{equation}
\frac{{b_{m}}}{{a_{m}}}=\lambda _{m}\eta ^{\nu }.
\end{equation}%
Using this result into Eq. (\ref{coef_rel_2_DO}), one finds
\begin{equation}
\frac{\Gamma (d_{+})}{\Gamma (d_{-})}=-\frac{1}{\lambda _{m}\gamma ^{\nu }}%
\frac{\Gamma (1+\nu )}{\Gamma (1-\nu )}.  \label{energy_BG_DO}
\end{equation}
Equation (\ref{energy_BG_DO}) implicitly determines the bound state energy
for the system for different values of the self-adjoint extension parameter.
Two limiting values for the self-adjoint extension parameter deserves some
attention. For $\lambda _{m}=0$, when the $\delta $ interaction is absent,
only the regular solution contributes for the bound state wave function. In
the other side, for $\lambda _{m}=\infty $, only the irregular solution
contributes for the bound state wave function. For all other values of the
self-adjoint extension parameter, both regular and irregular solutions
contributes for the bound state wave function. The energy for the limiting
values are obtained from the poles of gamma function, namely,
\begin{equation}
\left\{
\begin{array}{lll}
d_{+}=-n & \mbox{for }\lambda _{m}=0, & \mbox{(regular solution)}, \\
d_{-}=-n & \mbox{for }\lambda _{m}=\infty , & \mbox{(irregular solution)},%
\end{array}%
\right.  \label{gammapoles}
\end{equation}%
with $n$ a nonnegative integer, $n=0,1,2,\ldots $. By manipulation of Eq. (%
\ref{gammapoles}), we obtain
\begin{eqnarray}
E^{2}-M^{2} &=&2\sqrt{\frac{e^{2}B_{0}^{2}}{4}+2a\left( E+M\right) }  \notag
\label{energy_ABDO} \\
&\times &\left[ 2n+1\pm \sqrt{\left( m-e\phi \right) ^{2}+2b\left(
E+M\right) }\right]  \notag \\
&+&e^{2}B_{0}\phi -meB_{0}-esB_{0},\;\;S=V,
\end{eqnarray}%
\begin{eqnarray}
E^{2}-M^{2} &=&2\sqrt{\frac{e^{2}B_{0}^{2}}{4}+2a\left( E-M\right) }  \notag
\\
&\times &\left[ 2n+1\pm \sqrt{\left( m+s-e\phi \right) ^{2}+2b\left(
E-M\right) }\right]  \notag \\
&+&e^{2}B_{0}\phi -\left( m+s\right) eB_{0}-esB_{0},\;\;S=-V.
\end{eqnarray}
As a illustration the profiles of the energy under the exact spin symmetry
limit ($S=V$) as a function of the magnetic field $B_{0}$ and with spin
projection parameter $s=1$ and $s=-1$ are shown in Fig.~\ref{gra1} and Fig.~%
\ref{gra2}, respectively. From Fig.~\ref{gra1} and Fig.~\ref{gra2} we can
note that the ground state $n=0$ correspond to the lowest energies, as it
should be for particle energy levels.

In particular, it should be noted that for the case when $\nu \geq 1$ or
when the $\delta $ interaction is absent, only the regular solution
contributes for the bound state wave function ($b_{m}=0$), and the energy is
given by Eq. (\ref{energy_ABDO}) using the plus sign. The unnormalized bound
state wave functions for our problem are
\begin{eqnarray}
f_{m}(r)&=&N_{f}\left( A_{1}\right) ^{\pm \frac{1}{4}\sqrt{A_{2}}} r^{\pm
\sqrt{A_{2}}}e^{-\frac{1}{2}\sqrt{A_{1}}r^{2}} \notag \\
&\times& \,F\left( -n,1\pm \sqrt{A_{2}},\sqrt{A_{1}}r^{2}\right) ,
\end{eqnarray}%
for $S=V$, and%
\begin{eqnarray}
g_{m}(r)&=&N_{g}\left( B_{1}\right) ^{\pm \frac{1}{4}\sqrt{B_{2}}} r^{\pm
\sqrt{B_{2}}}e^{-\frac{1}{2}\sqrt{B_{1}}r^{2}} \notag \\
&\times& \,F\left( -n,1\pm \sqrt{B_{2}},\sqrt{B_{1}}r^{2}\right) ,
\end{eqnarray}%
for $S=-V$, where $N_{f}$ and where $N_{g}$ are normalization constants, and%
\begin{eqnarray}
A_{1} &=&\frac{e^{2}B_{0}^{2}}{4}+2a\left( E+M\right) , \\
A_{2} &=&\left( m-e\phi \right) ^{2}+2b\left( E+M\right) , \\
B_{1} &=&\frac{e^{2}B_{0}^{2}}{4}+2a\left( E-M\right) , \\
B_{2} &=&\left( m+s-e\phi \right) ^{2}+2b\left( E-M\right)
\end{eqnarray}

The self-adjoint extension is related with the presence of the $\delta $
interaction. In this manner, the self-adjoint extension parameter must be
related with the $\delta $ interaction coupling constant $\phi s$. In fact,
as shown in Refs. \cite{PRD.2012.85.041701,AoP.2013.339.510} (see also Refs.
\cite{PLB.2013.719.467,JPG.2013.40.075007}), from the regularization of the $%
\delta $ interaction, it is possible to find such a relationship. Using the
regularization method, one obtains the following equation for the bound
state energy
\begin{equation}
\frac{\Gamma (d_{+})}{\Gamma (d_{-})}=-\frac{1}{r_{0}^{2\nu }}\left( \frac{%
\phi s+\alpha \nu }{\phi s-\alpha \nu }\right) \frac{1}{\gamma ^{\nu }}\frac{%
\Gamma \left( 1+\nu \right) }{\Gamma \left( 1-\nu \right) }.
\label{energy_KS_DO}
\end{equation}%
By comparing Eqs. (\ref{energy_BG_DO}) and (\ref{energy_KS_DO}), this
relation is found to be
\begin{equation}
\frac{1}{\lambda _{m}}=\frac{1}{r_{0}^{2\nu }}\left( \frac{\phi s+\alpha \nu
}{\phi s-\alpha \nu }\right)  \label{saeparam}
\end{equation}%
where $r_{0}$ is a very small radius which comes from the $\delta $
regularization \cite{PRD.2012.85.041701,AoP.2013.339.510}. The result of Eq.
(\ref{saeparam}) provides an explicit formula for the self-adjoint extension
parameter $\lambda _{m}$. We have, therefore, derived the most important
quantities for the system without any arbitrary parameter coming from the
self-adjoint extension method.

\section{Nonrelativisitic limit}

\label{sec7}

Let us now examine the nonrelativistic limit of Eq. (\ref{diracB}) by
setting $E=M+\mathcal{E}$, with $M\gg \mathcal{E}$, for both cases $S=V$ and
$S=-V$. After applying this limit, we find
\begin{equation}
H\psi =2M\mathcal{E}\psi ,  \label{pauliosnr}
\end{equation}%
where%
\begin{equation}
H=\left( \mathbf{p}-e\mathbf{A}\right) ^{2}-es\boldsymbol{\sigma }\cdot
\mathbf{B}+2M\left[ S\left( r\right) +V\left( r\right) \right] .  \label{hlm}
\end{equation}%
Using the ansatz of Eq. (\ref{ansatz}) in Eq. (\ref{pauliosnr}), again, we
get the equation for $f_{m}\left( r\right) $ (for $S=V$)
\begin{gather}
\left[ -\frac{d^{2}}{dr^{2}}-\frac{1}{r}\frac{d}{dr}+\frac{\bar{\nu}^{2}}{%
r^{2}}+\bar{\eta}^{2}\text{ }r^{2}-es\phi \frac{\delta (r)}{r}\right]
f_{m}\left( r\right)  \notag \\
-\mathcal{E}f_{m}\left( r\right) =0.
\end{gather}%
where%
\begin{eqnarray}
\bar{\nu}^{2} &=&\left( m-e\phi \right) ^{2}+4Mb, \\
\bar{\eta}^{2} &=&\frac{e^{2}B_{0}^{2}}{4}+4Ma, \\
\bar{k}^{2} &=&meB_{0}-e^{2}B_{0}\phi +esB_{0}+2M\mathcal{E}.
\end{eqnarray}%
On the other hand, for $S=-V$, the term involving the potential is now
identically zero. The resulting equation is given by
\begin{gather*}
\left[ -\frac{d^{2}}{dr^{2}}-\frac{1}{r}\frac{d}{dr}+\frac{\breve{\nu}^{2}}{%
r^{2}}+\breve{\eta}^{2}\text{ }r^{2}-es\phi \frac{\delta (r)}{r}\right]
g_{m}\left( r\right) \\
-\mathcal{E}g_{m}\left( r\right) =0,
\end{gather*}%
where
\begin{eqnarray}
\breve{\nu}^{2} &=&\left( m+s-e\phi \right) ^{2}, \\
\breve{\eta}^{2} &=&\frac{e^{2}B_{0}^{2}}{4}, \\
\breve{k}^{2} &=&eB_{0}\left( m+s\right) -e^{2}B_{0}\phi +esB_{0}+2M\mathcal{%
E},
\end{eqnarray}%
In order to determine the energy spectrum, we use the same technique above.
Performing the same steps as for the relativistic case, one obtains the
energy levels%
\begin{eqnarray}
\mathcal{E} &=&\frac{1}{M}\sqrt{\frac{e^{2}B_{0}^{2}}{4}+4Ma}\left[ 2n+1\pm
\sqrt{\left( m-e\phi \right) ^{2}+4Mb}\right]  \notag \\
&+&\frac{1}{2M}\left[ e^{2}B_{0}\phi -\left( m+s\right) eB_{0}-esB_{0}\right]
,\;\;S=V,  \label{esv1}
\end{eqnarray}%
\begin{eqnarray}
\mathcal{E} &=&\frac{1}{2M}eB_{0}\left( 2n+1\pm \left\vert m-e\phi
\right\vert \right)  \notag \\
&+&\frac{1}{2M}\left[ e^{2}B_{0}\phi -\left( m+s\right) eB_{0}-esB_{0}\right]
,\;\;S=-V,  \label{esv2}
\end{eqnarray}%
The corresponding wave functions are given by
\begin{align}
& f_{m}(r)=\left( \frac{e^{2}B_{0}^{2}}{4}+4Ma\right) ^{\pm \frac{1}{4}\sqrt{%
\left( m-e\phi \right) ^{2}+4Mb}}  \notag \\
& \times \,r^{\pm \sqrt{\left( m-e\phi \right) ^{2}+4Mb}}e^{-\frac{1}{2}%
\sqrt{\frac{e^{2}B_{0}^{2}}{4}+4Ma}r^{2}}  \notag \\
& \times \,F\left( -n,1\pm \sqrt{\left( m-e\phi \right) ^{2}+4Mb},\sqrt{%
\frac{e^{2}B_{0}^{2}}{4}+4Ma}r^{2}\right),
\end{align}
for $S =V$,
\begin{align}
f_{m}(r)& =\left[ \frac{e^{2}B_{0}^{2}}{4}\right] ^{\pm \frac{1}{4}%
\left\vert m-e\phi \right\vert }r^{\pm \left\vert m-e\phi \right\vert }e^{-%
\frac{1}{4}eB_{0}r^{2}}\,  \notag \\
&\times \,F\left( -n,1\pm \left\vert m-e\phi \right\vert ,\frac{1}{2}%
eB_{0}r^{2}\right),
\end{align}
for $S=-V$.

\section{Conclusions}

\label{sec8}

In this paper, we have studied the relativistic quantum dynamics of a
spin-1/2 charged particle with minimal, vector and scalar couplings. The
minimal coupling was chosen as being one which leads to the spin-1/2
Aharonov-Bohm effect. In a first attempt, we have solved the equation of
first order Dirac. We verified that there are isolated solutions for the
system for some special cases. These solutions depend on the values assumed
by the spin projection parameter $s$, as well as on the choice of the scalar
$S\left(r\right)$ and vector $V\left( r\right)$ potential functions.

In contrast to what was addressed in the literature, we have considered the
correct quadratic form of the Dirac equation with minimal, vector and scalar
couplings. As we have mentioned before, in approach of Ref. \cite%
{AoP.2014.341.153}, the authors have not taken into account the term that
depends explicitly on the spin in the Pauli equation of motion. We have
revisited the dynamics of the system in detail and show that the correct
approach should involve the spin element. In this order, we have derived the
Pauli equation and studied the dynamics of the system taking into account
exact symmetry spin and pseudospin limits. Because of the equation of motion
includes a $\delta $ function, we have used the self-adjoint extension
method to specify the proper boundary conditions at the origin. The
analytical solutions of the model allow a calculation of expressions for the
energy eigenvalues and wave functions for both symmetry limits. We verify
that the presence of the spin element in the equation of motion introduces a
correction in the expressions for the bound state energy and wave functions,
a fact that does not occur in Ref. \cite{AoP.2014.341.153}.

\section*{Acknowledgments}

This work was supported by the CNPq, Brazil, Grants No. 482015/2013-6
(Universal), No 455719/2014-4 (Universal), No. 306068/2013-3 (PQ),
304105/2014-7 (PQ) and FAPEMA, Brazil, Grants No. 00845/13 (Universal), No.
01852/14 (PRONEM).

\end{document}